\documentclass{ws-procs9x6}

\begin{document}

\title{STANDARD SUPERSYMMETRY FROM A PLANCK-SCALE STATISTICAL THEORY}

\author{ROLAND E. ALLEN$^*$, ZORAWAR WADIASINGH, and SEIICHIROU YOKOO}

\address{Physics Department, Texas A\&M University,  
College Station, Texas 77843, U.S.A.\\
$^*$email: allen@tamu.edu}

\begin{abstract}
We outline three new ideas in a program to obtain standard physics, 
including standard supersymmetry, from a Planck-scale statistical theory: 
(1) The initial spin 1/2 bosonic fields are transformed to spin 0 fields
together with their auxiliary fields. (2) Time is defined by the
progression of 3-geometries, just as originally proposed by DeWitt.
(3) The initial (D-1)-dimensional ``path integral'' is converted 
from Euclidean to Lorentzian form by transformation of the fields 
in the integrand.
\end{abstract}

\bodymatter

\vspace{0.5cm}

In earlier work it was shown that a fundamental statistical
theory (at the Planck scale) can lead to many features of standard
physics\cite{allen-1997,allen-2002,allen-2003}.
In some respects, however, the results had nonstandard features which appear
to present difficulties. For example, the primitive supersymmetry of the
earlier papers is quite different from the standard formulation of
supersymmetry which works so admirably in both  protecting the masses of
Higgs fields from quadratic divergences and predicting coupling constant
unification at high energy. Also, the fact that the theory was originally
formulated in Euclidean time seems physically unsatisfactory for reasons
mentioned below. Here we introduce some refinements in the theory which
eliminate these two problems. The ideas in the following sections 
respectively grew out of discussions of the first author with 
Seiichirou Yokoo (on the 
transformation of spin 1/2 to spin 0 fields) and Zorawar Wadiasingh 
(on the transformation of the path integral from Euclidean to Lorentzian 
form).

\vspace{-0.3cm}
\section{Transformation of Original Spin 1/2 Fields Yields Standard 
Supersymmetry}

       In Refs. 2 and 3, the action for a fundamental bosonic field 
       was found to have the form 

\begin{eqnarray}
S_{b}=\int d^{4}x\,\psi _{b}^{\dagger }\,i\sigma ^{\mu }\partial _{\mu }\psi
_{b}
\end{eqnarray}
at energies that are far below the Planck energy $m_{P}$ (with $\hbar =c=1)$
and in a locally inertial coordinate system. This is the conventional 
form of the action
for fermions, described by 2-component Weyl spinors, but it is highly
unconventional for bosons, because a boson described by $\psi _{b}$ would
have spin 1/2. We can, however, transform from the original 2-component
field $\psi _{b}$ to two 1-component complex fields $\phi $ and $F$ by
writing 
\begin{eqnarray}
\psi _{b}\left( x\right) &=&\psi ^{+}\left( x\right) +\psi ^{-}\left( x\right) \\
\psi ^{+}\left( \vec{x},t\right) &=& \sum_{\vec{p},\omega }\phi \left( \vec{p}
,\omega \right) u^{+}\left( \vec{p}\right) e^{i\vec{p}\cdot \vec{x}
}e^{-i\omega t}\left( \omega +\left\vert \vec{p}\right\vert \right)
^{1/2} \\
\psi ^{-}\left( \vec{x},t\right) &=& \sum_{\vec{p},\omega
}F\left( \vec{p},\omega \right) u^{-}\left( \vec{p}\right) e^{i\vec{p}\cdot 
\vec{x}}e^{-i\omega t}\left( \omega +\left\vert \vec{p}\right\vert \right)
^{-1/2}
\end{eqnarray}
with 
\begin{eqnarray}
\vec{\sigma}\cdot \vec{p}\,u^{+}\left( \vec{p}\right) =+\left\vert \vec{p}
\right\vert u^{+}\left( \vec{p}\right) \quad ,\quad \vec{\sigma}\cdot \vec{p}
\,u^{-}\left( \vec{p}\right) =-\left\vert \vec{p}\right\vert u^{-}\left( 
\vec{p}\right) 
\end{eqnarray}
\begin{eqnarray}
\phi \left( \vec{p},\omega \right) =\int d^{4}x\,\phi \left( \vec{x}
,t\right) e^{-i\vec{p}\cdot \vec{x}}e^{i\omega t}\; ,\; F\left( \vec{p}
,\omega \right) =\int d^{4}x\,F\left( \vec{x},t\right) e^{-i\vec{p}\cdot 
\vec{x}}e^{i\omega t}\, . \nonumber
\end{eqnarray}
Substitution then gives 
\begin{eqnarray}
S_{b} &=& V^{-1} \sum_{\vec{p},\omega }\,\left[ \phi ^{\ast }
\left( \vec{p},\omega
\right) \left( \omega ^{2}-\left\vert \vec{p}\right\vert ^{2}\right) \phi
\left( \vec{p},\omega \right) +F^{\ast }\left( \vec{p},\omega \right)
F\left( \vec{p},\omega \right) \right]  \\
&=&\int d^{4}x\, \left[ - \partial ^{\mu }\phi ^{\ast }\left( x\right) \partial
_{\mu }\phi \left( x\right) +F^{\ast }\left( x\right) F\left( x\right) 
\right] 
\end{eqnarray}
where $\partial ^{\mu }=\eta ^{\mu \nu }\partial _{\nu }$, $\eta ^{\mu \nu
}=diag\left( -1,1,1,1\right) $, and $V$ is a 4-dimensional
normalization volume. This is, of course, precisely the action for
a massless scalar boson field $\phi $ and its auxiliary field $F$.

With the fermionic action left in its original form, we
now have the standard supersymmetric action for each pair of susy partners: 
\begin{equation}
S_{fb}=\int d^{4}x\,\left[ \psi _{f}^{\dagger }\,i\sigma ^{\mu }\partial
_{\mu }\psi _{f} - \partial ^{\mu }\phi ^{\ast }\left( x\right) \partial _{\mu
}\phi \left( x\right) +F^{\ast }\left( x\right) F\left( x\right) \right] \,.
\end{equation}

There is a major point that will be discussed at length elsewhere, in a more
complete treatment of the present theory: The above transformation works
only for $\omega +\left\vert \vec{p}\right\vert \ge 0$, since otherwise
the sign of the integrand would be reversed. However, a stable vacuum already
requires $\omega \ge 0$, so we must define time for would-be
negative-frequency fields in such a way that this condition is satisfied.

\section{Time is Defined by Progression of $3$-Geometries in External Space}

In our earlier work, the time coordinate $x^{0}$ was initially defined in 
exactly the same way as each spatial coordinate $x^{k}$, 
so $x^{0}$ was initially a Euclidean variable. For 
reasons given in the following section, however, this does not seem to be as
physically reasonable as a picture in which time is Lorentzian when it
is first defined. In this section, therefore, we move to a 
new picture in which the initial ``path integral'' $Z_{E}$ still has the Euclidean
form \noindent 
\begin{equation}
Z_{E}=\int \mathcal{D}\left( \mathrm{Re\,}\phi\right) \,\mathcal{D}
\left( \mathrm{Im\,}\phi\right) \,\,e^{-S} \; ,\; S=\int
d^{D-1}x\,\phi^{\ast }\left( \vec{x}\right) \,A\,\phi\left( \vec{x}
\right) 
\end{equation}
but there is \textit{initially no time}. We are then confronted with the 
well-known situation in canonical quantum gravity\cite{DeWitt}, where the
``wavefunction of the universe'' is a 
functional of only 3-geometries, with no time dependence. Roughly speaking,
cosmological time is then defined by the cosmic scale factor $R$ (except
that there can be different branches for the state of the
universe, corresponding to, e.g., expansion and contraction, as well
as different initial conditions). More precisely, the progression of 
time is locally defined by the progression of local $3$-geometries.

An analogy is a stationary state for a proton with
coordinates $\vec{X}$ passing a hydrogen atom with coordinates 
$\vec{x}$. The time-independent Schr\"{o}dinger equation can be written
\begin{equation}
\left( -\frac{\hbar ^{2}}{2m_{p}}\nabla _{p}^{2}+H_{e}\right) \Psi \left( 
\vec{X}\right) \psi \left( \vec{x},\vec{X}\right) =E\,\Psi \left( \vec{X}
\right) \psi \left( \vec{x},\vec{X}\right) 
\end{equation}
with $\Psi $ required to satisfy
\begin{equation}
-\frac{\hbar ^{2}}{2m_{p}}\nabla _{p}^{2}\,\Psi \left( \vec{X}\right)
=E\,\Psi \left( \vec{X}\right) .
\end{equation}
Then the equation for $\psi $ is 
\begin{equation}
\left( -\frac{\hbar ^{2}}{m_{p}}\Psi^{-1}\vec{\nabla}_{p}\Psi \cdot \vec{\nabla}_{p}-
\frac{\hbar ^{2}}{2m_{p}}\nabla _{p}^{2}+H_{e}\right) \psi \left( \vec{x},
\vec{X}\right) =0.
\end{equation}
The first term involves a local proton velocity 
\begin{equation}
\vec{v}_{p}=\hbar \,\vec{\nabla}_{p}\theta /m_{p}\qquad ,\qquad \Psi
=\left\vert \Psi \right\vert e^{i\theta }.
\end{equation}
For a state in which the proton is moving rapidly, with 
\begin{equation}
\Psi =\Psi _{0}e^{i\vec{P}\cdot \vec{X}/\hbar },
\end{equation}
and in which $\left( \hbar ^{2}/2m_{p}\right) \nabla _{p}^{2}\,\psi $ is
relatively small, we obtain 
\begin{equation}
i\hbar \frac{\partial }{\partial t}\psi \left( \vec{x},t\right) =H_{e}\,\psi
\left( \vec{x},t\right) \qquad ,\qquad \frac{\partial }{\partial t}\equiv 
\frac{\vec{P}}{m_{p}}\cdot \vec{\nabla}_{p}\,.
\end{equation}
\bigskip \noindent One then has an ``internal
time'' defined within a stationary state\cite{Reading}. Similarly, one
can define time as a progression of $3$-geometries, just as proposed 40
years ago by DeWitt, whose formulation of canonical quantum gravity 
(following the classical canonical decomposition of Arnowitt, Deser, and
Misner, and the work of Dirac, Wheeler, and others) involves the local 
canonical momentum operator 
\begin{equation}
\pi ^{kl}\left( \vec{x}\right) =-\,i\frac{\delta }{\delta g_{kl}
\left( \vec{x}\right) },
\end{equation}
which corresponds to the proton momentum operator $-\,i\,\hbar\vec{\nabla}_{p}$ 
in the 
analogy above. After introducing the $3$-dimensional metric tensor in the
way described in Refs. 1-3, and the gravitational action in a way that
will be described in a more complete treatment, we move from the original 
path-integral quantization to canonical quantization, with a state 
\begin{equation}
\Psi _{\mathrm{total}}=\Psi _{\mathrm{gravity}}\left[ g_{kl}\left( \vec{x}
\right) \right] \;\Psi _{\mathrm{otherfields}}\left[ \phi _{\mathrm{
otherfields}}\left( \vec{x}\right),g_{kl}\left( \vec{x}\right) \right] 
\end{equation}
and time is defined essentially in the same way as in the analogy.

\section{Transformation of 3-Dimensional ``Path Integral'' Changes Euclidean 
Factor $\mathbf{e^{-S}}$ to Lorentzian Factor $\mathbf{e^{iS}}$}

A Euclidean path integral with the form of (9), but
with time included, is formally transformed into a Lorentzian path integral 
\begin{equation}
Z_{L}^{D}=\int \mathcal{D}\left( \mathrm{Re}\, \phi _{L}\right) 
\,\mathcal{D}\left( 
\mathrm{Im} \, \phi _{L}\right) \,e^{iS_{L}^{D}}\qquad ,\qquad S_{L}^{D}=\int
d^{D}x\,\mathcal{L}_{L}
\end{equation}
through an inverse Wick rotation 
$x_{E}^{0}=t_{E}\rightarrow ix_{L}^{0}=it_{L}$.
$S_{L}^{D}$ has the usual form of a classical action, and it leads to the usual
description of quantized fields via path-integral quantization. In other
words, the standard equations of physics follow from $S_{L}^{D}$, and are
therefore formulated in Lorentzian time. The Euclidean formulation, in
either coordinate or momentum space, is ordinarily regarded as a mere
mathematical tool which can simplify calculations and make them better
defined.

Hawking, on the other hand, has suggested that Euclidean spacetime may
actually be more fundamental than Lorentzian spacetime. In his well-known 
popular book, he says\cite{Hawking1} ``So maybe
what we call imaginary time is really more basic, and what we call real is
just an idea that we invent to help us describe what we think the universe
is like.'' And in a more technical paper he states\cite{Hawking2} ``In fact
one could take the attitude that quantum theory and indeed the whole of
physics is really defined in the Euclidean region and that it is simply a
consequence of our perception that we interpret it in the Lorentzian
regime.'' 

However, there is a fundamental problem with this point of view, because the
factor $\,e^{iS_{L}^{D}}$ in the Lorentzian formulation results in interference
effects, whereas the factor $\,e^{-S_{E}^{D}}$ in the Euclidean formulation does
not. Also, a formal transformation from $t_{E}$ to $t_{L}$ mixes all of 
the supposedly more fundamental Euclidean times in the single
Lorentzian time that we actually experience. Finally, it appears
difficult to formulate a mathematically well-founded and physically
well-motivated transformation of a general path integral from Euclidean to
Lorentzian spacetime.

Here we adopt a very different point of view: (1) Nature is
fundamentally statistical, essentially as proposed in Refs. 1-3, but the 
initial path integral (or partition function) does not contain 
the time as a fundamental coordinate. Instead time is defined by the
local 3-space geometry (or more generally, (D-1)-space geometry). (2) 
It is, however, still necessary to transform from the Euclidean
form (9), with $e^{-S}$, to the Lorentzian form (18), with
$e^{iS}$ (but also with no time coordinate, so that $D \rightarrow
D-1$ in (18)), and this is our goal in the present section.

Consider a single complex scalar field $\phi$ with a $
3$-dimensional ``Euclidean path integral'' 
\begin{equation}
Z_{E}=\int \mathcal{D}\left( \mathrm{Re\,}\phi\right) \,\mathcal{D}
\left( \mathrm{Im\,}\phi\right) \,\,e^{-S}\quad ,\quad S=\int
d^{3}x\,\phi^{\ast }\left( \vec{x}\right) \,A\,\phi\left( \vec{x}
\right) .
\end{equation}
In a discrete picture, the operator $A$ is replaced by a matrix with
elements $A\left( \vec{x},\vec{x}^{\prime }\right) $: 
\begin{equation}
S=\sum_{x,x^{\prime }}\phi^{\ast }\left( \vec{x}\right) \,A\left( \vec{x
},\vec{x}^{\prime }\right) \,\phi\left( \vec{x}^{\prime }\right) .
\end{equation}
$A$ can be diagonalized to $A\left( \vec{k},\vec{k}^{\prime
}\right) =a\left( \vec{k}\right) \delta _{\vec{k},\vec{k}^{\prime }}$.
Then 
\begin{eqnarray}
Z_{E} \equiv \left[ \prod_{\vec{x}}\int_{-\infty }^{\,\infty }d\left( \mathrm{Re}
\,\phi\left( \vec{x}\right) \right) \int_{-\infty }^{\,\infty }d\left( 
\mathrm{Im}\,\phi\left( \vec{x}\right) \right) \,\right] \exp \left(
-\sum_{\vec{x},\vec{x}^{\prime }}\phi^{\ast }\left( \vec{x}\right)
\,A\left( \vec{x},\vec{x}^{\prime }\right) \,\phi\left( \vec{x}^{\prime
}\right) \right) \nonumber 
\end{eqnarray}
becomes\cite{Peskin}
\begin{equation}
Z_{E}=\left[ \prod_{\vec{k}}\int_{-\infty }^{\,\infty }d \, \mathrm{Re}\,\phi
\left( \vec{k}\right) \int_{-\infty }^{\,\infty }d \, \mathrm{
Im}\,\phi\left( \vec{k} \right) \,\right] \exp \left( -\sum_{
\vec{k}}\phi^{\ast }\left( \vec{k}\right) \,a\left( \vec{k}\right) \phi
\,\left( \vec{k}\right) \right) .
\end{equation}
The Gaussian integrals over $\mathrm{Re}\,\phi\left( \vec{k}
\right) $ and $\mathrm{Im}\,\phi\left( \vec{k}\right) $ may be
evaluated as usual at each $\vec{k}$ to give 
\begin{equation}
Z_{E}=\prod_{\vec{k}}\frac{\pi }{a\left( \vec{k}\right) }=\frac{\prod_{\vec{k}
}\pi }{\mathrm{det\,}A}.
\end{equation}

Here, and in the earlier papers, two representations of the path integral are
taken to be physically equivalent if they give the same result for all
operators $A$ (including those which produce zero except for arbitrarily
restricted regions of space and sets of fields). 
For example, we might define a path integral $Z^{\prime }$ with 
fields $\phi^{\prime }$ and $\bar{\phi}^{\prime }$ which are
treated as independent and which each vary along the real axis. It is 
then appropriate to include the formal Jabobian, 
with a value of $1/2$, which would correspond to a  
transformation from $\mathrm{Re}\,\phi$ and $\mathrm{Im}\,\phi$ to 
$\phi^{\prime }=\mathrm{Re}\,\phi+i\,\mathrm{Im}\,\phi$ and $
\bar{\phi}^{\prime } =i\left( \mathrm{Re}\,\phi-i\,
\mathrm{Im}\,\phi\right) $. Since
\begin{eqnarray}
Z^{\prime } &\equiv &\left[ \prod_{\vec{k}}\frac{1}{2}\int_{-\infty
}^{\,\infty }d\left( \phi^{\prime }\left( \vec{k}\right) \right)
\int_{-\infty }^{\,\infty }\,d\left( \bar{\phi}^{\prime }\left( \vec{k}
\right) \right) \right] \exp \left( \sum_{\vec{k}}i\,\bar{\phi}
^{\prime }\left( \vec{k}\right) \,a\left( \vec{k}\right) \phi
^{\prime }\left( \vec{k}\right) \right) \nonumber \\
&=&\prod_{\vec{k}}\frac{1}{2}\int_{-\infty }^{\,\infty }d\left( \phi
^{\prime }\left( \vec{k}\right) \right) \int_{-\infty }^{\,\infty
}\,d\left( \bar{\phi}^{\prime }\left( \vec{k}\right) \right) \exp \left(
i\,\bar{\phi}^{\prime }\left( \vec{k}\right) \,a\left( \vec{k}\right)
\;\phi^{\prime }\left( \vec{k}\right) \right)  \nonumber \\
&=&\prod_{\vec{k}}\frac{1}{2a\left( \vec{k}\right) }\int_{-\infty
}^{\,\infty }d\left( a\left( \vec{k}\right) \;\phi^{\prime }\left( \vec{
k}\right) \right) \,2\pi \,\delta \left( \,a\left( \vec{k}\right) \;\phi
^{\prime }\left( \vec{k}\right) \right) \\
&=&\prod_{\vec{k}}\frac{\pi }{a\left( \vec{k}\right) } \\
&=&Z_{E}
\end{eqnarray}
for any operator $A$, we regard $Z_{E}$ and $Z'$ as being physically equivalent.

Now let us define a ``Lorentzian path integral'' $Z_{L}$ by
\begin{eqnarray} 
Z_{L}&=&\int \mathcal{D}\left( \mathrm{Re\,}\phi\right) \,\mathcal{D}
\left( \mathrm{Im\,}\phi\right) \,\,e^{iS}\ \\
&\equiv&\left[ \prod_{\vec{x}}\, \frac{1}{i} \, \int_{-\infty }^{\,\infty }d\left( 
\mathrm{Re}
\,\phi\left( \vec{x}\right) \right) \int_{-\infty }^{\,\infty }d\left( 
\mathrm{Im}\,\phi\left( \vec{x}\right) \right) \,\right] \exp \left(
i\sum_{\vec{x},\vec{x}^{\prime }}\phi^{\ast }\left( \vec{x}\right)
\,A\left( \vec{x},\vec{x}^{\prime }\right) \,\phi\left( \vec{x}^{\prime
}\right) \right) . \nonumber 
\end{eqnarray}
Diagonalization of $A$ gives
\begin{eqnarray}
Z_{L}&=&\left[ \prod_{\vec{k}} \frac{1}{i} \int_{-\infty }^{\,\infty }
d \, \mathrm{Re}\,\phi
\left( \vec{k}\right) \int_{-\infty }^{\,\infty }d \, \mathrm{
Im}\,\phi\left( \vec{k} \right) \,\right] \exp \left( i\sum_{
\vec{k}}\phi^{\ast }\left( \vec{k}\right) \,a\left( \vec{k}\right) \phi
\,\left( \vec{k} \right) \right) \nonumber \\
&=&\prod_{\vec{k}} \frac{1}{i} \, \frac{i\, \pi }{a\left( \vec{k}\right) }\\
&=&Z_{E} \; .
\end{eqnarray}
Then $Z_{E}$ can be replaced by $Z_{L}$, which involves the original 
operator $A$ and the original 
spatial coordinates $\vec{x}$, but a different form for the integrand. 
This replacement is possible because time is introduced only after 
$Z$ is in Lorentzian form.

The transformation from $Z_{E}$ to $Z_{L}$ can be regarded as a 
transformation of the fields in the integrand, with the lines along which
$\mathrm{Re\,}\phi$  and $\mathrm{Im\,}\phi$ are integrated
each being rotated by $45^\circ$ in the complex plane\cite{Swanson}.

\section{Outline of Broad Program: From a Planck-Scale Statistical 
Theory to Standard Physics with Supersymmetry}

The ideas above are part of a broad program to obtain 
standard physics, including supersymmetry, from a description at the
Planck scale which is purely statistical. The major steps in the
complete program are as follows:

(1) The fundamental statistical picture gives a $D-1$ ``Euclidean 
action'' for bosons only (and with no time yet): 
\begin{equation}
Z_{b}^{D-1}=\int \mathcal{D}\left( \mathrm{Re\,}\phi \right) \,\mathcal{D}\left( 
\mathrm{Im\,}\phi \right) \,e^{-S_{b}}\quad ,\quad S_{b}=\int d^{D-1}x\,
\mathcal{L}_{b}^{D-1} \,.
\end{equation}

(2) Random fluctuations then give a ``Euclidean action'' with bosons, 
fermions, and a primitive supersymmetry:
\begin{equation}
Z_{E}^{D-1}=\int \mathcal{D}\left( \mathrm{Re\,}\phi \right) \,\mathcal{D}\left( 
\mathrm{Im\,}\phi \right) \,\mathcal{D}\left( \mathrm{Re\,}\psi \right) \,
\mathcal{D}\left( \mathrm{Im\,}\psi \right) \,e^{-S}\quad ,\quad S=\int
d^{D-1}x\,\mathcal{L}^{D-1} \, .
\end{equation}

(3) Transformation of the integrand in the ``path integral'' 
changes the ``Euclidean factor'' $e^{-S}$ to the ``Lorentzian factor'' $e^{iS}$:
\begin{equation}
Z_{L}^{D-1}=\int \mathcal{D}\left( \mathrm{Re\,}\phi \right) \,\mathcal{D}\left( 
\mathrm{Im\,}\phi \right) \,\mathcal{D}\left( \mathrm{Re\,}\psi \right) \,
\mathcal{D}\left( \mathrm{Im\,}\psi \right) \,e^{iS} \quad ,\quad S=\int
d^{D-1}x\,\mathcal{L}^{D-1} \, .
\end{equation}

(4) The 3-dimensional gravitational metric tensor $g_{kl}$ and 
$SO(N)$ gauge fields  $A_{k}$ (and their 
initial, primitive supersymmetric partners) result from rotations 
of the vacuum state vector, 
in both $3 $-dimensional external space and $D-4$ dimensional internal space.

(5) Time is defined by the progression of $3$-geometries in external space.

(6) The Einstein-Hilbert action for the gravitational field (as well as the
cosmological constant), the Maxwell-Yang-Mills action for the gauge fields, and
the analogous terms for the gaugino and gravitino fields are assumed to 
arise from a response of the vacuum that is analogous to the diamagnetic 
response of electrons.

(7) The gravitational field is approximately quantized via first a 
path-integral formulation and then the canonical formulation of Ref. 4.

(8) Heisenberg equations of motion are then obtained for all fields.

(9) Transformation of the initial spin 1/2 bosonic fields, followed by
definition of standard gaugino and gravitino fields, gives standard 
supersymmetry.

(10) One finally obtains an effective action which is the same as that of
standard physics with supersymmetry, except that particle masses, 
Yukawa couplings, and self-interactions are assumed
to arise from supersymmetry breaking and radiative corrections.

A more complete treatment will be given in a much longer paper.

\end{document}